%% file: main.tex
\documentclass[sigconf,screen]{acmart}
\renewcommand\footnotetextcopyrightpermission[1]{} %
\setcopyright{none}
\acmDOI{}
\acmISBN{}
\acmPrice{}

\usepackage[small,compact]{titlesec}
\usepackage[T1]{fontenc}
\usepackage[english]{babel}
\usepackage[normalem]{ulem}
\usepackage{url}
\usepackage{amsmath}
\usepackage{xspace}
\usepackage[skip=2pt]{caption}
\usepackage{subfig}

\usepackage{etoolbox,xpatch}
\usepackage[frozencache,cachedir=minted-cache]{minted}
\setminted{fontsize=\footnotesize}
\setminted{fontfamily=cmtt}
\makeatletter %
\AtBeginEnvironment{minted}{\dontdofcolorbox}
\def\dontdofcolorbox{\renewcommand\fcolorbox[4][]{##4}}
\xpatchcmd{\inputminted}{\minted@fvset}{\minted@fvset\dontdofcolorbox}{}{}
\xpatchcmd{\mintinline}{\minted@fvset}{\minted@fvset\dontdofcolorbox}{}{}
\makeatother

\usepackage{adjustbox}
\usepackage{xcolor}
\usepackage{multirow}
\usepackage{lipsum}
\usepackage{nicefrac}
\usepackage{enumitem}
\usepackage{outlines}
\usepackage{makecell}
\usepackage{cleveref}
\usepackage{circledsteps}
\usepackage{algorithm,algorithmicx}
\usepackage[noend]{algpseudocode}
\algrenewcommand\algorithmicindent{1em}

\newcommand{\codefont}[1]{\textit{#1}}

\newcommand{\captionfonts}{\bf\small}
\makeatletter  %
\long\def\@makecaption#1#2{%
	\vskip\abovecaptionskip
	\sbox\@tempboxa{{\captionfonts #1: #2}}%
	\ifdim \wd\@tempboxa >\hsize
	{\captionfonts #1: #2\par}
	\else
	\hbox to\hsize{\hfil\box\@tempboxa\hfil}%
	\fi
	\vskip\belowcaptionskip}
\makeatother

\expandafter\def\expandafter\normalsize\expandafter{%
    \normalsize
    \setlength\abovedisplayskip{3pt}
    \setlength\belowdisplayskip{3pt}
    \setlength\abovedisplayshortskip{3pt}
    \setlength\belowdisplayshortskip{3pt}
}

\newcommand{\squishlist}{
  \begin{list}{$\bullet$}{
    \setlength{\itemsep}{0pt}       \setlength{\parsep}{3pt}
    \setlength{\topsep}{3pt}        \setlength{\partopsep}{0pt}
    \setlength{\leftmargin}{1em}    \setlength{\labelwidth}{1em}
    \setlength{\labelsep}{0.5em} } }
\newcommand{\squishend}{\end{list}}

\def\compactify{\itemsep=2pt \topsep=2pt \partopsep=1pt \parsep=1pt \leftmargin=1.2em}
\let\latexusecounter=\usecounter

\begin{document}

\title[Symphony]{Symphony: Optimized DNN Model Serving using Deferred Batch Scheduling}
\date{}

\author{Lequn Chen}
\affiliation{\institution{University of Washington}}
\author{Weixin Deng}
\affiliation{\institution{University of Washington}}
\author{Anirudh Canumalla}
\affiliation{\institution{University of Washington}}
\author{Yu Xin}
\affiliation{\institution{University of Washington}}
\author{Danyang Zhuo}
\affiliation{\institution{Duke University}}
\author{Matthai Philipose}
\affiliation{\institution{Microsoft}}
\author{Arvind Krishnamurthy}
\affiliation{\institution{University of Washington}}

\input{tex/00-abstract}

\maketitle\pagestyle{plain}
\input{tex/10-intro}
\input{tex/20-back}

\input{tex/30-schedule}

\input{tex/40-design}

\input{tex/45-partition}

\input{tex/70-eval}

\input{tex/90-conclusion}

\newpage

\bibliographystyle{ACM-Reference-Format}
\bibliography{references}

\input{tex/99-appendix}

\end{document}

%% file: tex/00-abstract.tex
\begin{abstract}

Having large batch sizes is one of the most critical aspects of increasing the accelerator efficiency and the performance of DNN model inference. However, existing model serving systems cannot achieve adequate batch sizes while meeting latency objectives as these systems eagerly dispatch requests to accelerators to minimize the accelerator idle time. We propose Symphony, a DNN serving system that explores deferred batch scheduling to optimize system efficiency and throughput.
Further, unlike other prior systems, Symphony's GPU usage is \emph{load-proportional}: it consolidates workloads on the appropriate number of GPUs and works smoothly with cluster auto-scaling tools.
Symphony consists of two core design points. First, Symphony defines a \emph{schedulable window} in which a batch of inference requests can be dispatched. This window is computed in order to improve accelerator efficiency while meeting the request's SLO. 
Second, Symphony implements a scalable, low-latency, fine-grained coordination scheme across accelerators to dispatch and execute requests in the schedulable window.
Through extensive scheduler-only benchmarks, we demonstrate that Symphony can schedule millions of requests per second and coordinate thousands of GPUs while also enabling robust autoscaling that adapts to workload changes.
Symphony outperforms prior systems by achieving 5x higher goodput when given the same number of GPUs and 60\% reduction in GPUs when given the same workload.

\end{abstract}

%% file: tex/10-intro.tex
\section{Introduction}

We consider the setting of a cloud-scale inference service that supports the scalable execution of a high-volume inference workload involving many deep neural network (DNN) models on a cluster of accelerators (e.g., GPUs). Such an inference service is increasingly needed to consolidate the ML needs of cloud services within a datacenter. An effective inference service has to provide not only high throughput for a diverse set of models but also results within tight latency bounds that are appropriate for user-facing cloud services.

In many aspects, a cloud-scale inference service does not differ much from the traditional cloud service model; it would have to balance incoming requests across backends, adapt to workload changes (i.e., autoscale), and achieve high efficiency without compromising on the desired latency service-level objectives (SLOs).
However, scheduling DNN inference requests presents a unique challenge, distinct from traditional request scheduling, primarily due to the \textit{batching effect}. DNNs use linear algebra operators, and batched execution of these operators improves the utilization efficiency of the accelerators. While existing serving systems~\cite{Clipper,TFServing,nexus,clockwork,shepherd} batch requests, they struggle to achieve optimal batch sizes. The root cause is that the existing serving systems always try to eagerly dispatch a batch of requests whenever there is an idle accelerator to minimize device idle time. This leads to smaller batches and reduced batch efficiency and system throughput. In addition, eager dispatching also creates challenges for cluster horizontal autoscaling because all GPUs are always busy regardless of the workload.

In this paper, we explore a new approach for request scheduling, called \textit{deferred batch scheduling}. Deferring requests enables the system to:
(1) accumulate a larger number of requests, thereby increasing the batch size and throughput,
and (2) consolidate the usage of GPUs so that the number of GPUs used is proportional to the load.

Designing a DNN inference system using this idea raises two key challenges. First, \textit{how long can we afford to defer a batch?} Inference requests have tight SLOs, typically around 20--50 ms. Adding an excessive delay may lead to requests violating their SLOs. The second challenge lies in \textit{facilitating scalable, low-latency coordination among accelerators}. This requires the scheduler to efficiently track the status of each accelerator and dispatch requests in a manner that optimizes system performance and efficiency.

To this end, we build Symphony, a DNN inference system with two key design aspects. First, Symphony defines a \textit{schedulable window} for a request. Symphony dispatches requests only in this window. This window is computed to enhance the batch size and, at the same time, avoid SLO violations. The size of the window depends on the batching effect (i.e., the GPU throughput vs. batch size curve). For example, when the batching effect is strong, Symphony prefers smaller and later schedulable windows to allow the batch to accumulate more requests before being scheduled. 

The second aspect is a series of system optimizations to enable scalable, low-latency coordination so that the scheduler can always dispatch requests to the GPU in the schedulable window. These optimizations include
(1) separation of request batching and model-GPU matchmaking,
(2) low time-complexity and multicore scalable scheduling algorithm,
(3) low-latency and predictable control and data planes.

Our evaluation includes extensive comparisons of Symphony with other serving systems and shows that Symphony can improve serving system performance across a wide range of workloads. 
Symphony further differentiates itself from other systems by exhibiting ideal \textit{load-proportional} autoscaling properties that enable effective adaptation to workload changes. This is because Symphony can keep batch sizes consistently large, while previous systems reduced batch sizes when load is reduced, thus wasting GPU resources. We demonstrate that Symphony can handle 12 million requests per second, coordinate thousands of GPUs, and adapt to workload changes. Compared to state-of-the-art model serving systems, Symphony provides up to 5x gain in goodput when given the same number of GPUs and up to 60\% cut in the number of GPUs when serving the same workload. 

\textit{This work does not raise any ethical issue.}

%% file: tex/20-back.tex
\section{Background on Inference Serving Systems}

\subsection{The Batching Effect in DNN Serving}

We consider scheduling DNN inferences on accelerators, such as GPUs. DNNs are networks of linear algebra operations called {\em layers} or {\em kernels}, where the networks are typically referred to as {\em models}. 
Batching groups input matrices into higher-dimensional ones before applying custom ``batched'' implementations of the GPU kernels. It is one of the most, if not the most, important techniques to improve GPU utilization. This is because accelerators, like GPUs, are designed for highly parallel operations. Batching uses GPU resources (including both processing cores and GPU memory) more effectively and amortizes GPU memory I/O
and GPU kernel launch time for individual inference requests.
Previous work has found that DNN model execution is highly predictable~\cite{clockwork} and the latency profile can be fit with a linear function with high fidelity~\cite{nexus,shepherd}:
\begin{equation*}
\label{eq:batch}
\ell(b) = \alpha b + \beta
\end{equation*}
where $b$ is batch size, $\beta$ is the fixed cost of invoking a model on a batch of requests, and $\alpha$ is the cost of each additional task in the batch. Large batches amortize the fixed cost $\beta$.

However, leveraging this batching effect is not simple. 
Requests must be satisfied within latency bounds set by the application. In the context of datacenter applications (e.g., serving ads or real-time detection tasks), the latency SLOs are a few tens of milliseconds and constrain the extent to which requests can be batched. 
To quantify a model serving system's performance under latency SLOs, we define its \emph{goodput} as the highest aggregate throughput over all models such that the p99 tail latency of each model is less than their respective latency SLO.

\begin{figure}[t]
\centering
\includegraphics[width=\linewidth]{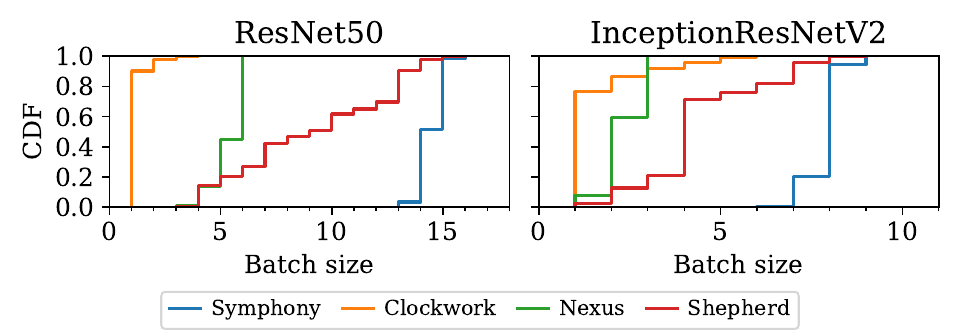}
\caption{Batch size distribution}
\label{fig:stagger}
\end{figure}

\subsection{Batch Scheduling in Existing Serving Systems}

How do existing DNN serving systems make decisions on batching? There are currently three options. The first option is to schedule a batch of requests whenever a GPU is idle, i.e., \textit{eager batching}. The batch size is thus the number of requests accumulated before any GPU becomes idle. The rationale behind eager batching is to keep GPUs busy.
Nexus~\cite{nexus} and Clockwork~\cite{clockwork} use eager scheduling, and so does Shepherd~\cite{shepherd} as a default policy.
However, one key problem is that minimizing GPU idle time may lead to low GPU arithmetic intensities and, consequently, lower throughput.

The three systems have different ways of implementing eager scheduling.
Scheduling in Nexus happens in three components: the scheduler, frontends, and GPU backends. Every 10 seconds, the Nexus scheduler determines the models and their corresponding expected batch sizes for each GPU. Each Nexus frontend routes a DNN request to one of the GPU backends that run the requested model. A Nexus backend runs scheduler-assigned models using round-robin. When the actual batch size differs from the scheduler-assigned value, the Nexus backend either runs with the actual smaller batch size or drops excess requests.
Clockwork and Shepherd both use a centralized scheduler that monitors all incoming requests and dispatches batches to GPUs.
For an incoming request, Clockwork creates a batch candidate for every batch size and maintains these candidates for each GPU.
When a GPU becomes free, Clockwork dispatches the batch candidate whose latest executable moment is the earliest and invalidates related candidates for other GPUs. 
The latest executable moment, as defined by Clockwork, is the latest point at which initiating a batch's execution does not violate its deadline.
Shepherd only maintains one outstanding batch candidate for each model and dispatches the candidate with the biggest batch size when a GPU becomes free.

The second approach is \textit{eager batching with preemption}. Preemption is the idea of canceling the dispatched batch by terminating its execution to make room for a different batch. If the new batch has a larger batch size, this can improve system throughput. This is an enhancement to eager batching to avoid small batch sizes.
Sherpherd~\cite{shepherd} allows a running batch to be preempted by another if the new one is at least 3x the size. When the latency constraint is tight, and the batch size is small (e.g., smaller than 16), such preemption might occur less often. However, preemption can lead to wasted work because the in-flight batch is canceled. Canceling also has its overheads, so this is only worthwhile when the performance benefits can significantly outweigh the drawbacks. This results in systems still issuing small-size batches.

The third approach is \textit{time-out based batch scheduling}.
The serving system dispatches the batch to a GPU if the batch size has reached the maximum or the elapsed time since the first request arrival has exceeded the given timeout value. TensorFlow Serving~\cite{TFServing} uses the time-out-based approach and uses a constant timeout value and a constant maximum batch size value.
Tuning the two parameters introduces operation complexity, and they have to be changed when the workload changes.

To understand the batching quality of DNN model serving systems, we ran a single copy of ResNet50~\cite{resnet} (SLO 25 ms) and InceptionResNetV2~\cite{inceptionresnet} (SLO 70 ms) separately on 8 GPUs using Clockwork, Nexus, Shepherd, and Symphony. \Cref{fig:stagger} shows that the median batch sizes for the four systems are 1, 6, 9, 14 on ResNet50, and 1, 2, 4, 8 on InceptionResNetV2.
The batch sizes of prior systems are much smaller than those dispatched by Symphony.

\subsection{Batching and Auto-scaling}

\begin{figure}[t]
\centering
\includegraphics[width=\linewidth]{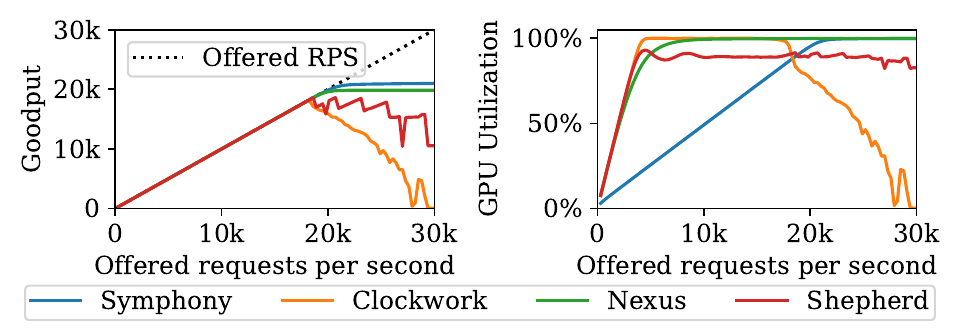}
\caption{(Left) Goodput. (Right) GPU Cluster Utilization}
\label{fig:stability-gpu-cycle}
\end{figure}

Having small batch sizes can also affect the effectiveness of auto-scaling.
When batch sizes are small, the GPUs are busy but inefficient. The autoscaling controller (e.g., Kubernetes) will mistakenly believe insufficient GPUs are allocated and ask for more GPUs. More GPUs means GPU idle events are triggered more frequently, subsequently making batch sizes even smaller. An ideal system should achieve a higher batch size and only ask for the required GPU resources it needs. Figure~\ref{fig:stability-gpu-cycle} shows the behavior of Clockwork, Nexus, and Shepherd when we scale the offered load from 0 to 30K requests per second. When the offered load is only 3K requests per second, all the GPUs are already fully utilized with small batch sizes in these existing model inference systems (Figure~\ref{fig:stability-gpu-cycle} (Right).) In comparison, our system, Symphony, only needs 20\% of the GPUs to serve the exact same workloads.

%% file: tex/30-schedule.tex
\section{Deferred Batch Scheduling}

The key idea of our design to improve the batching efficiency in DNN inference is Deferred Batch Scheduling, focusing on the optimal timing for dispatching a batch of requests to a GPU. In this section, we first discuss the notion of a \emph{schedulable window} and justify why we would want to schedule request batches in this window. We then present the details of our scheduling algorithms.

\subsection{Schedulable Window}
\label{sec:schedulable-window}

\begin{figure}[t]
\centering
\includegraphics[width=0.8\linewidth]{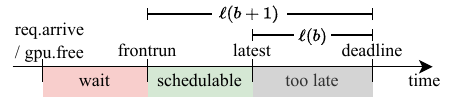}
\caption{Schedulable window}
\label{fig:schedulable-window}
\end{figure}

Let's consider the timing of a batch of requests in Figure~\ref{fig:schedulable-window}. Denote the batch size as $b$ and the deadline as $d$.
Now, we consider the time window for when such a batch can be dispatched to a GPU.

The latest moment is $d - \ell(b)$, because the execution of the batch takes $\ell(b)$ time.
Such a late binding guarantees the maximum batching efficiency while still meeting the deadline.
However, it might leave GPUs idle for too long.

We propose a new option to decide when a batch becomes schedulable.
We define \emph{frontrun} as $d - \ell(b+1)$, denoted as $f$.
We start to allow a batch to be dispatched to a GPU at \codefont{frontrun}.
Notice that if a new request arrived at time $t^{\star}$ between \codefont{frontrun} and \codefont{latest}, adding it to the batch would cause the batch to violate the deadline ($t^{\star} + \ell(b+1) > f + \ell(b+1) = d$).
Hence, dispatching the batch at \codefont{frontrun} maintains the same batching efficiency as dispatching at \codefont{latest}.
In addition, \codefont{frontrun} reduces GPU idle time compared to using \codefont{latest}.
We explicitly disallow dispatching a batch prior to \codefont{frontrun} because we want to accumulate a larger batch. Note that compared to the timeout-based approach, we do not need to specify two parameters.
As the batch size grows, the \codefont{frontrun} moment is pushed to an earlier time.
When the current time is between \codefont{frontrun} and \codefont{latest}, the batch is dispatched to an idle GPU.

\subsection{Scheduling Algorithm}

\begin{algorithm}[t]
\small
\caption{Scheduling Algorithm}\label{alg:frontrun-pseudocode}
\begin{algorithmic}[1]

\Procedure{UpdateCandidate}{$M$} \Comment{Model $M$}
    \State $B \gets \Call{GetBatch}{Q_M}$ \Comment{Max batch that can fit in deadline}
    \If{$|B| > 0$}
        \State $d \gets \mathrm{min}\left\{ r.\codefont{deadline} : r \in B \right\}$
        \State \codefont{exec} $\gets \mathrm{max}\left( \Call{now}{\null}, d - \ell_M(|B|+1) \right)$ \label{alg:line-assign-exec-at}
        \State \codefont{latest} $\gets d - \ell_M(|B|)$
        \State $c_M \gets \left( B, \codefont{exec}, \codefont{latest} \right)$
    \Else
        \;$c_M \gets \varnothing$
    \EndIf
\EndProcedure
\Procedure{Dispatch}{$M$, $G$} \Comment{Model $M$, GPU $G$}
    \State \Call{UpdateCandidate}{M} \Comment{Update exec}
    \State Send $c_M$ to $G$ for execution
    \State $G.\codefont{free} \gets c_M.\codefont{exec} + \ell_M(|c_M.B|)$
    \State Remove $c_M.B$ from $Q_M$
    \State \Call{UpdateCandidate}{M} \Comment{Prepare next batch}
\EndProcedure

\Procedure{OnNewRequest}{$r$} \Comment{Request $r$ for Model $M$}
    \State Enqueue $r$ to $Q_M$
    \State \Call{UpdateCandidate}{M}
\EndProcedure

\Procedure{OnModelTimer}{M} \Comment{Trigger at $c_M.\codefont{exec}$}
    \State $G^{\star} \gets \mathrm{argmin}_G \left\{ G.\codefont{id} : G.\codefont{free} < c_M.\codefont{exec} \right\}$
    \If{$G^{\star} \neq \varnothing$}
        \Call{Dispatch}{$M$, $G^{\star}$}
    \EndIf
\EndProcedure

\Procedure{OnGpuTimer}{G} \Comment{Trigger at $G.\codefont{free}$}
    \State $M^{\star} \gets \mathrm{argmin}_M \left\{ c_M.\codefont{latest} : c_M.\codefont{exec} < G.\codefont{free} < c_M.\codefont{latest} \right\}$
    \If{$M^{\star} \neq \varnothing$}
        \Call{Dispatch}{$M^{\star}$,$G$}
    \EndIf
\EndProcedure

\end{algorithmic}
\end{algorithm}

We present our scheduling algorithm, as listed in \Cref{alg:frontrun-pseudocode}.
We assume that the scheduler has global knowledge of incoming requests and GPU execution states.
\Cref{sec:design} will cover how to achieve it by a centralized scheduler design.

For each model $M$, the scheduler maintains two states: a request queue $Q_M$ and a candidate batch $c_M$.
For each GPU $G$, the scheduler keeps track of when it will be free.
The candidate batch is updated when a new request arrives or when a previous batch is dispatched to a GPU.
The candidate batch contains three components: a set of requests $B$, the desired time to run (\codefont{exec}),
and the \codefont{latest} time that the candidate remains valid.
The desired time to run is either the \codefont{frontrun} moment or the current time, whichever is later.

Subroutine \textsc{GetBatch} returns a maximum set of requests that can finish within the deadline.
Typically, the batch-gathering algorithm starts from the head of the request queue and then repeatedly adds the next request to the set if it can still meet the deadline~\cite{Clipper,shepherd}.
Alternatively, the batch-gathering algorithm can prematurely drop the head of the queue in order to maintain a larger target batch size~\cite{nexus}.
Our algorithm works well with both batch-gathering algorithms.

The Model-GPU matchmaking happens in the following two events.
Each model has a timer that triggers at $c_M.\codefont{exec}$.
When the candidate is updated, the old timer event is canceled, and the timer is reset to the new expiry.
When the model timer triggers, the scheduler tries to find a free GPU ($G.\codefont{free} < c_M.\codefont{exec}$) and then dispatches the finalized batch to the GPU.
If there are multiple GPUs available, the scheduler could pick an arbitrary one.
For example, we can pick the one with the smallest identifier (e.g., UUID).
This choice allows GPUs with bigger identifiers to remain completely idle when the system load is low.
If these GPUs remain idle for an extended period of time, cluster auto-scaling tools can release these GPUs.
On the other hand, if there is no free GPU at the moment, the candidate becomes schedulable and might be matched by a GPU later.

Each GPU has a timer that triggers
when it finishes execution.
When the GPU timer triggers, it considers candidates that are schedulable ($c_M.\codefont{exec} < G.\codefont{free}$) and still valid ($G.\codefont{free} < c_M.\codefont{latest}$).
When multiple such candidates are present, the algorithm picks the one with the \codefont{latest} moment that is closest, which prioritizes the urgency of a batch.

\subsection{Examples of batch scheduling}
\label{sec:schedule-example-stagger}

\begin{figure}[t]
\centering
\includegraphics[width=\linewidth]{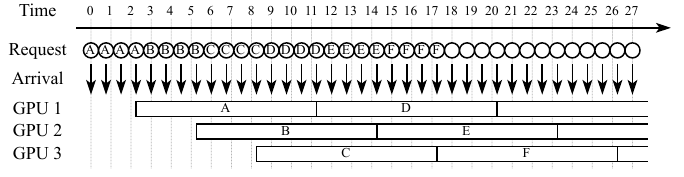}
\caption{Deferred scheduling forms a staggered execution. Circle: request; Letter: batch identifier.}
\label{fig:stagger-init}
\end{figure}

\begin{figure}[t]
\centering
\subfloat[Eager batch scheduling deteriorates]{\includegraphics[width=\linewidth]{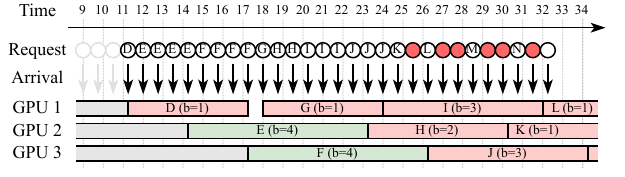}\label{fig:stagger-skip-eager}}\\
\subfloat[Deferred batch scheduling runs normally]{\includegraphics[width=\linewidth]{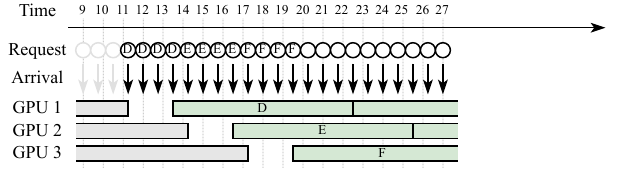}\label{fig:stagger-skip-frontrun}}
\caption{Reaction to three missing requests. Red circle: dropped request; Red/Green block: small/large batch.}
\label{fig:stagger-skip}
\end{figure}

We now use two examples to help understand the behavior of deferred batch scheduling and how it could achieve higher goodput and react to request rate fluctuation.
We consider a cluster of 3 GPUs and 1 model. The latency profile is $l(b)=b+5$ time units (i.e., $\alpha=1$ and $\beta=5$), and the SLO is 12. Consider a uniform arrival process with request $R_i$ arriving at $t=0.75 \cdot (i-1)$. Assume that all GPUs are free at the beginning. At $t=2.25$, $R_4$ arrives. The deadline of $R_1$ is at $t=12$. The frontrun moment is $t=12-\ell(5)=2$ and the latest is $t=12-\ell(4)=3$. Therefore, the first batch, including the first four requests, is dispatched to a GPU. The rest of the scheduling can be reasoned about in a similar way. \Cref{fig:stagger-init} shows the execution trace of the deferred batch scheduling. The GPUs form a \emph{staggered execution} pattern. In addition, the deferred batch scheduling is able to maintain this pattern.

Note that the staggered execution achieves the highest batch efficiency as it bounds the worst-case queueing delay while dispatching uniformly large batches. We can obtain an upper bound on model goodput using the following analysis.
Denote the number of GPUs as $N$, batch size as $b$, request rate as $\lambda$.
In the staggered execution pattern, the worst queuing delay for a request is $\nicefrac{\ell(b)}{N}$.
The latency constraint and the throughput requirement give the following two equations:
\begin{align}
    \label{eqn:batch-size}
    \left(1 + \nicefrac{1}{N}\right) \cdot \ell(b) &\leq \mathrm{SLO} \\
    \label{eqn:throughput}
    N \cdot \nicefrac{b}{\ell(b)} &\geq \lambda
\end{align}
Combining the two equations, we can solve for $b$ and $N$, which stands for the optimal configuration. We use this to analyze the scheduling quality in \Cref{sec:eval-scheduling-quality}.

Next, we make a small change to the request arrival process and examine how it affects the staggered execution pattern under different scheduling algorithms. We skip $R_{13}$, $R_{14}$, $R_{15}$ while $R_{16}$ and later requests remain unchanged. We study the reactions to the three missing requests in deferred and eager scheduling.
If the system uses eager scheduling starting from $R_{16}$, the throughput will deteriorate, as depicted in \Cref{fig:stagger-skip-eager}. At $t=11.25$, $R_{16}$ arrives and GPU 1 finishes the previous batch. Eager scheduling immediately dispatches $R_{16}$ to GPU 1. Then, GPU 2 and GPU 3 each dispatch a batch of size 4 when they finish the last batch. At $t=18$, $R_{25}$ arrives, and GPU 1 runs with batch size 1 again. At $t=23.25$, GPU 2 finishes, and the system has accumulated 7 requests ($R_{26}$--$R_{32}$). However, the system can only run with batch size 2 because the deadline of $R_{26}$ is at $t=30.75$. This decreasing batch size eventually causes the system to drop $R_{35}$, $R_{37}$, $R_{38}$, and so on.
On the other hand, \Cref{fig:stagger-skip-frontrun} shows that deferred batch scheduling lets GPUs idle for a short period and regains the staggered execution pattern again.

\subsection{Empirically validating deferred batch scheduling}

In this subsection, we present empirical benchmarks against eager batch scheduling and timeout-based batch scheduling to justify our design of deferred batch scheduling.
Timeout-based batch scheduling can be implemented by changing Line~\ref{alg:line-assign-exec-at} of \Cref{alg:frontrun-pseudocode} to $\codefont{exec} \gets \mathrm{max}(\textsc{now}(), a+k)$ where the earliest request arrival time $a = \mathrm{min}\left\{ r.\codefont{arrival} : r \in B \right\}$ and $k$ is the constant timeout value.
In particular, $k=0$ is equivalent to eager scheduling.
Goodput is found by a binary search over sending a fixed request rate.
This subsection assumes that all models are equally popular.

\subsubsection{Case Studies}

\begin{figure}[t]
\centering
\subfloat[Compared with Eager]{\includegraphics[width=0.5\linewidth]{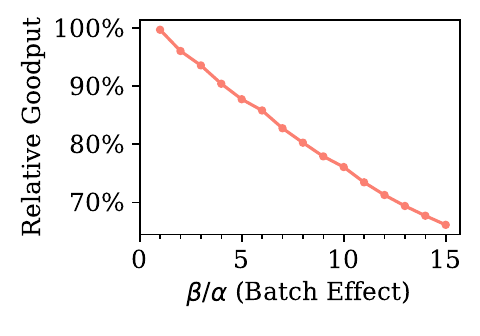}\label{fig:frontrun-case-study-beta}}
\subfloat[Compared with Timeout]{\includegraphics[width=0.5\linewidth]{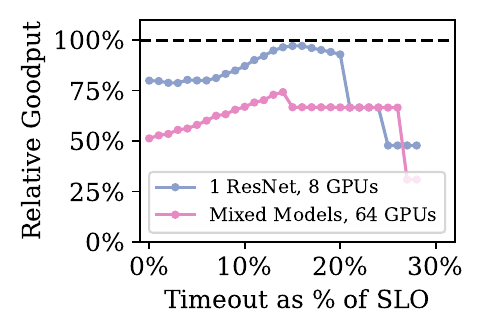}\label{fig:frontrun-case-study-timeout}}
\caption{Case study of Deferred Batch Scheduling}
\label{fig:frontrun-case-study}
\end{figure}

First, we use case studies to deepen our understanding of deferred batch scheduling.
In \Cref{fig:frontrun-case-study-beta}, we consider the impact of a model's batch characteristic and compare eager scheduling with deferred scheduling. The figure plots the goodput achieved by eager scheduling as a percentage of deferred scheduling's goodput.
We use a synthetic latency profile where $\alpha$ is 1 ms and $\beta$ varies from 1 ms to 15 ms, and we set latency SLO to $2\ell(8)$.
The simulated cluster has 32 GPUs and 10 models of the same latency profile. The request arrival pattern follows Poisson distribution.
The figure shows that when $\beta/\alpha=1$, deferred batch scheduling and eager scheduling have nearly the same goodput.
When the model's batching effect is stronger, deferred batch scheduling has a bigger advantage.

In \Cref{fig:frontrun-case-study-timeout}, we compare deferred batch scheduling to different timeout settings of timeout-based batch scheduling.
We consider two simulation setups: single ResNet50 with 50 ms latency SLO on an 8-GPU cluster and 37 mixed models with various SLO (\Cref{tab:full-latency-profiles-a100}) on a 64-GPU cluster.
We set the timeout value of each model to be a varying percentage of the latency SLO.
\Cref{fig:frontrun-case-study-timeout} shows the relative goodput compared to the deferred batch scheduling.
As the timeout value increases, the timeout-based approach gets closer to the deferred one because a larger batch can be accumulated with a longer timeout.
However, when the timeout value is too big, we can see that goodput of the timeout-based approach drops because it might let GPUs idle for too long.
In the single-model case, the optimal timeout value can reach the same goodput as the deferred batch scheduling.
However, in the presence of multiple models, the goodput of the timeout-based batch scheduling is much lower than that of the deferred batch scheduling.
A potential way to improve multi-model performance with a timeout-based approach is to carefully tune the timeout value for each model, which will introduce significant operational overhead, especially when the set of models served by the system changes.

\subsubsection{Synthetic Workload}

\begin{table}[t]
{\small
\begin{tabular}{|l|l|}\hline
Model           & \makecell[l]{DenseNet121, InceptionV3, \\ResNet50V2, VGG16, Xception, Bert} \\\hline
\#Models        & 8, 16, 24, 32, 48, 64 \\\hline
\#GPUs:\#Models & 1.0, 1.5, 2.0, 2.5, 3.0, 3.5, 4.0                                                       \\\hline
SLO (ms)        & 20, 25, 30, 40, 50                                                                      \\\hline
Burstiness      & $\Gamma(.1)$,  $\Gamma(.2)$,  $\Gamma(.3)$,  $\Gamma(.5)$,  $\Gamma(.7)$,  $\Gamma(1.)$ \\\hline
\end{tabular}
}
\caption{Configurations for synthetic workloads}
\label{tab:beta-lambda-exp-setup}
\end{table}

\begin{figure*}[htb]
\centering
\includegraphics[width=\linewidth]{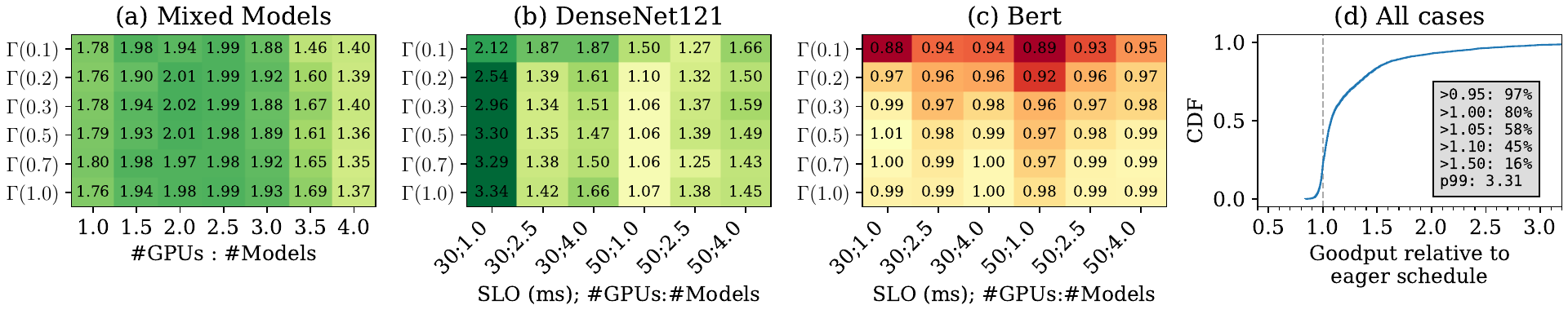}
\caption{Goodput of the deferred batch scheduling relative to the goodput of the eager batch scheduling}
\label{fig:beta-lambda}
\end{figure*}

We evaluate 5880 different synthetic workloads.
We consider homogeneous-model setups where all models have the same latency profile and mixed model setups. Table~\ref{tab:beta-lambda-exp-setup} lists options for each configuration dimension. Models listed in Table~\ref{tab:beta-lambda-exp-setup} are ordered by descending batching effect ($\beta/\alpha$ ranging from 9.7 to 0.02). A mixed-model setup contains 35 different models in a single run. Details about these models are in the Appendix \ref{app:modelzoo}.
The arrival process follows the Gamma distribution, with the shape varying from 0.1 to 1.0. A smaller Gamma shape value represents a burstier request pattern. Notably, $\Gamma(1.0)$ is mathematically equivalent to the Exponential distribution, meaning the request rate follows the Poisson distribution.

Figure~\ref{fig:beta-lambda}(a) shows that the deferred batch scheduling works very well when serving a mixture of models, gaining 35\% to 102\% higher goodput.
Figure~\ref{fig:beta-lambda}(b) examines a few cases with DenseNet121, which has a strong batching effect. Our deferred batch scheduling shows 34\% to 234\% goodput increase under tighter latency SLO constraint (30ms). For looser latency constraints (50ms), the advantage is less prominent.
Figure~\ref{fig:beta-lambda}(c) examines a few cases with Bert. Bert is a model with a weak batching effect, which should favor eager scheduling. Our deferred batch scheduling achieves a goodput similar to eager scheduling in these extreme cases.
Figure~\ref{fig:beta-lambda}(d) plots all cases. For almost all cases, the deferred batch scheduling is no worse than (>0.95) the eager scheduling. The deferred batch scheduling achieves 50\% higher goodput for 16\% cases and more than double in extreme cases.

\subsection{Auto-scaling Support}

Batching not only has an effect on efficiency but also on how the system can perform auto-scaling.
We desire a certain form of schedule stability that will aid us in performing auto-scaling in response to workload properties. In particular, we want our scheduler to exhibit the following two properties:

\squishlist
\item \textbf{Goodput Stability:} Assume that the peak goodput of a cluster for a given workload is $p$. If we were to issue an offered load $o$ higher than $p$, we would like our scheduler to have a bad rate comparable to $(o-p)/o$. Otherwise, in overloaded conditions, the system exhibits a goodput lower than the peak goodput, indicating a suboptimal congestion response.

\item \textbf{Load-Proportional GPU Usage:} Given a peak goodput $p$, if we were to issue an offered load $o$ lower than $p$, then we would like our scheduling system to have an average GPU idle time fraction comparable to $(p-o)/p$. (GPU idle time fraction is defined as the fraction of time the GPU spends idling without executing any tasks.)
\squishend

We refer to these two properties together as a desired \emph{flat-top} behavior.
The flat-top behavior is desirable because a sudden burst in the aggregate load should not cause an undue increase in bad rate. More importantly, the scheduling system can then monitor the system performance signals of bad rate and GPU idle time to determine whether it needs to allocate or deallocate resources from the cluster.
For example,

\noindent \textbf{Allocate GPUs:} If the bad rate $r$ is above a threshold, the autoscaler requests $N \cdot r/(1-r) $ additional GPUs, where $N$ is the current number of cluster GPUs.

\noindent \textbf{Deallocate GPUs:} If the GPU idle time fraction is $f$, then it deallocates $N \cdot f$ GPUs.

It is easy to achieve flat-top behavior when serving tasks that do not exhibit batch amortization benefits. However, when there are varying degrees of batch amortization across different candidates with different batch sizes, it is hard to get a robust signal for the autoscaling capability. For example, if a scheduler were to use suboptimal batch sizes, then the GPU idle time would underestimate the true overprovisioning available in the system. Similarly, in overloaded conditions, the bad rate could provide an overestimate of additional resources required to achieve parity.

\Cref{fig:stability-gpu-cycle} is a measurement of the same workload on the four systems. Symphony and Nexus achieve goodput stability. 
Shepherd and Clockwork exhibit a substantial loss in goodput as we increase the offered load beyond their capabilities.
Symphony's GPU usage is load proportional due to deferred batch scheduling, whereas all three previous systems are eager to occupy all GPUs before reaching their peak goodputs, creating challenges for cluster auto-scaling. We will describe the design of Symphony in \Cref{sec:design}, and we will discuss this experiment in more detail in \Cref{sec:eval-auto-scaling}.

%% file: tex/40-design.tex
\section{System Design and Implementation}
\label{sec:design}

In this section, we describe the system design of Symphony.

\subsection{System architecture}

\begin{figure}[ht]
\centering
\includegraphics[width=\linewidth]{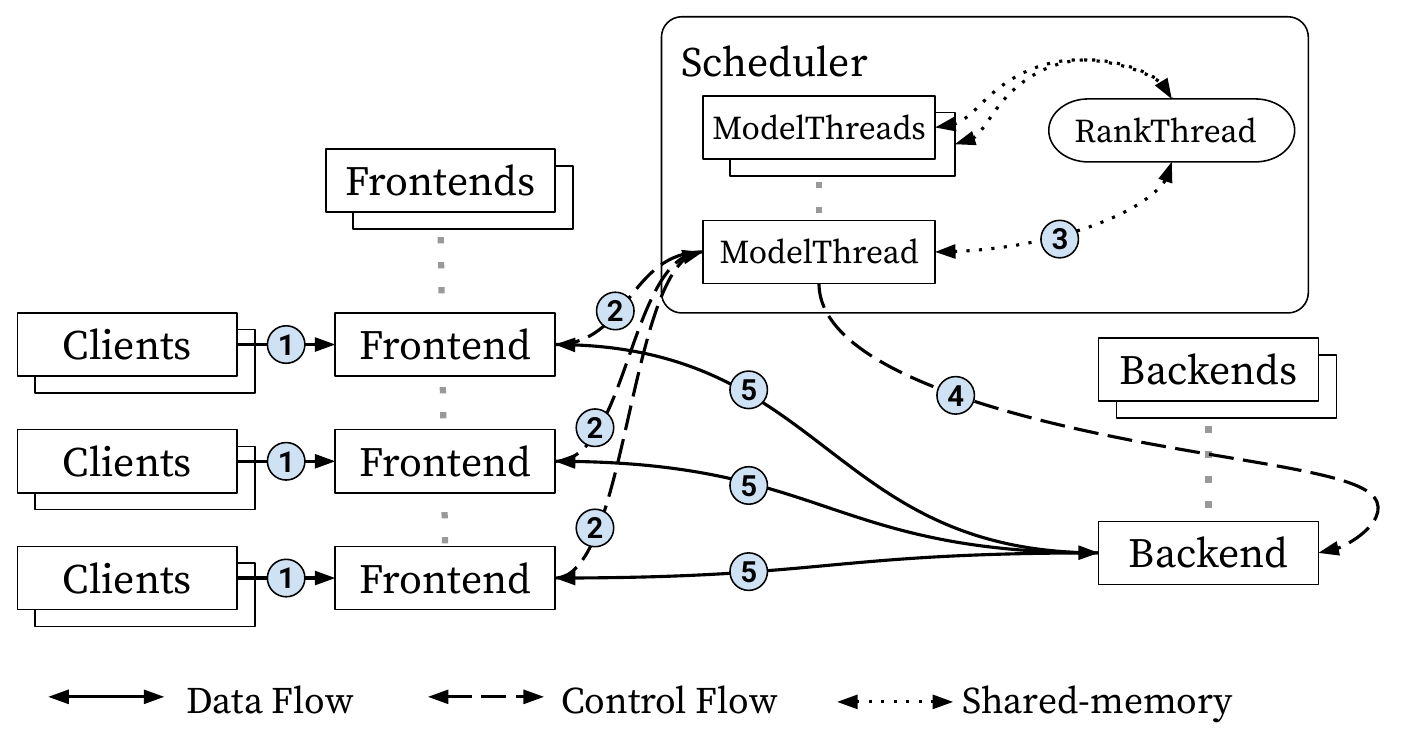}
\caption{Symphony Architecture Design}
\label{fig:arch-diagram}
\end{figure}

Now we outline the overall system architecture and how the different system components work together (depicted in Figure~\ref{fig:arch-diagram}). As with other model-serving systems, frontends represent nodes that run application logic and invoke model inferences. Multiple frontends running the same or different applications can invoke inference tasks for the same model. The model inferences are performed in batches on a cluster of backends, each equipped with GPUs.

Frontends accept requests from clients~\Circled[]{1}, then communicate tasks and their deadlines to the scheduler~\Circled{2}. Tasks are concisely represented using unique task IDs. The scheduler determines the batching of inference tasks and identifies the backend that will execute a given batch~\Circled{3}. Once the scheduler has sent the metadata of the batch to the chosen backend~\Circled{4}, the backend directly pulls the input data for the inference tasks from the frontends~\Circled{5}. When the batch finishes, the backend pushes the outputs to the frontends.

Similar to previous work~\cite{shepherd}, backends are grouped into different sub-clusters, where GPUs in the same sub-cluster have the same set of models loaded at a given time.
Periodically, over epochs of a few minutes, Symphony performs the following tasks: allocating and loading models on sub-clusters and auto-scaling to react to workload changes.

Our deferred batch scheduling algorithm requires the scheduler to have global knowledge of the cluster, including request information and GPU work states. Therefore, we use this centralized scheduler design.
Centralized scheduling brings three benefits. First, the scheduler can dynamically schedule batches of inference tasks to any one of the available backends to obtain statistical multiplexing benefits, wherein a bursting model can use resources left unused by models operating with lower-than-expected loads; in other words, it can balance out short-term variations in loads across different models. Second, a centralized scheduler can evenly distribute loads across successive dispatches of a model's batches to various GPU backends. Third, a centralized scheduler can reduce the queueing delay for batches by accumulating just one active batch for each model type and funneling it to the next available GPU. This reduced queueing delay translates to being able to execute larger batches within the latency SLO and an increase in GPU efficiency.

\subsection{Scalability of the centralized scheduler}
\label{sec:design-scalability}

The centralized scheduler is architected to process millions of requests per second and manage thousands of GPUs.

At the scheduler, when we update a batch candidate, we do not need to access other models' information. On the other hand, GPU availability is global information that all models need to access when trying to bind a candidate to a GPU. For multi-core scalability, we separate the scheduler into the following two different kinds of entities.

A \textsc{ModelThread} accepts incoming requests to a particular model. It accesses only model-local information and updates the candidate. The candidate is then sent to \textsc{RankThread}.

\emph{The} \textsc{RankThread} organizes the global information: GPU free time, each model's timer, and each GPU's timer. Model-GPU matchmaking is triggered by the timers and is handled in the \textsc{RankThread}. If matchmaking succeeds, \textsc{RankThread} sends a ``GPU Granted'' message to the matched \textsc{ModelThread} and marks the GPU as unavailable.

When the corresponding \textsc{ModelThread} receives the ``GPU Granted'' message, it updates the candidate and sends out the finalized batch to the GPU backend immediately. It also informs the \textsc{RankThread} about when the GPU will become available. Then it registers with the \textsc{RankThread} a new candidate, which contains tasks for the next batch.

With this design, the centralized scheduler can utilize multiple independently executing \textsc{ModelThread}s spread across multiple CPU cores. The \textsc{RankThread}, however, is shared by all \textsc{ModelThread}s and represents the bottleneck in the system. Note that although a \textsc{ModelThread} needs to keep up with the request line rate, the \textsc{RankThread} only needs to be fast enough to keep up with the start and the finish events of GPU execution. Since the GPU execution is batched, usually with batch sizes larger than 10, the process rate for \textsc{RankThread} is an order of magnitude lower.

Besides, we have engineered the system to have a limited amount of ranking logic in the \textsc{RankThread}, and the single \textsc{RankThread} is able to support dozens of \textsc{ModelThread}s.
With the help of advanced data structures~\cite{splay-tree}, the algorithm time complexity on new requests and on batch completion are both $O(\log M + \log G)$ where $M$ is the number of models and $G$ is the number of GPUs.

Combining these techniques, Symphony's centralized scheduler can scale to millions of requests per second on a modern multi-core server.
For reference, we put an extended pseudo-code of \Cref{alg:frontrun-pseudocode} in \Cref{appendix:pseudocode}, which captures the scalability design.

\subsection{Fast and predictable networking}

Deferred batch scheduling and the centralized scheduler put pressure on networking.
(1) Metadata for every request and every batch has to traverse through the scheduler~\Circled{2}\Circled{4}.
(2) The batching algorithm needs to account for the worst latency of the network delay.
(3) A backend cannot fetch input data from frontends until it has received the batch metadata from the scheduler.

Therefore, it is critical to have a low-latency, high-bandwidth, and predictable connection among the scheduler, frontends, and backends. We use one-sided RDMA READs initiated from the backend to fetch inference inputs from the frontends. We use two-sided RDMA for control plane messages. Networking behavior is studied in detail in Section~\ref{sec:eval-tail-latency}.

Besides reading inputs, preprocessing inputs (e.g., decoding images) also takes a significant amount of CPU time. In our design, we choose to do preprocessing at frontends for two reasons. One is that the cluster can easily add CPU computation power by adding more frontends. More importantly, Symphony can then overlap the preprocessing time with the request's queueing time instead of introducing an additional delay at the backends before GPU computations.

%% file: tex/45-partition.tex
\subsection{Sub-cluster Partitioning}

Every a few minutes, we invoke a global partitioning algorithm for solving the following resource allocation problem. Given a cluster and a set of models to be served from the cluster, we subdivide the cluster into multiple sub-clusters and an associated set of models for each sub-cluster. We then ensure that each backend in a sub-cluster is pre-loaded with all models associated with the sub-cluster. This ensures that any of the backends in a sub-cluster can execute any associated models and increases the ability to consolidate the loads across models. Note that a single centralized scheduler is still responsible for scheduling all sub-clusters.

The primary constraint considered in partitioning a set of models across multiple sub-clusters is the total memory size of the models associated with a sub-cluster. This constraint arises because each backend in a sub-cluster needs to load all of the sub-cluster's associated models into bounded GPU memory. This lets the dispatcher send a model batch to any of the GPUs associated with the sub-cluster and simplifies the centralized scheduling problem. In addition to this primary constraint, we let operators impose an additional optional bound on the combined request rate that a sub-cluster can handle, as a configuration parameter that reflects network capacities and other resource constraints.

To address these constraints, we decide to distribute the workload at the granularity of models. In particular, we partition the set of all models in the system into disjoint sets of models, with each associated with a sub-cluster and managed by a corresponding dispatcher thread on the centralized scheduler. The benefit of such a design is that each dispatcher thread can function independently, so no communications between dispatcher threads are needed.  %

Because the set of models and their request rates can change over time, we also want to minimize the disruption of such changes on the partition because backends won't be able to process requests during the loading and unloading of models.
In order to find a particular partition, we formulated the following mixed integer linear programming (MILP) model. Appendix~\ref{app:partition} provides details on the MILP and an evaluation of its effectiveness.

%% file: tex/70-eval.tex
\section{Evaluation}

We implemented Symphony with 10k lines of C++ code. We use TensorFlow v2.5.0 as the engine to run DNNs. All models are profiled with all different batch sizes to obtain actual execution latency.

We ran most evaluations in a 9-machine cluster. Each machine has two sockets of Intel Xeon E5-2690 v4 CPU (28 physical cores in total) and 64GB memory. The cluster is interconnected by 56Gbps Infiniband with Mellanox ConnectX-3 network interface card. 8 of the machines are equipped with one NVIDIA GeForce 1080Ti GPU on each machine, and the other machine without a GPU is used as the scheduler.

The same cluster is also used to emulate bigger clusters of more GPUs and faster GPU cards. Since the execution time of DNNs on GPU is highly predictable~\cite{clockwork}, we emulate the execution by simply introducing a delay at the backend. The introduced delay times are based on model profiles from 1080Ti and A100. We implemented the emulation mechanism for Symphony, Clockwork, Nexus, and Shepherd.

Clockwork uses TVM as the engine to run DNNs. We found that the TVM version shipped with Clockwork runs significantly slower than TensorFlow. To minimize the impact of different DNN execution engines, we always use emulated GPUs based on the profiling results on TensorFlow in Clockwork's end-to-end experiments. Due to the limitation of TVM, Clockwork only supports power-of-two batch sizes. To make comparisons fairer, we added support for arbitrary batch sizes to Clockwork.
Clockwork routes all input data through their centralized controller, creating a bandwidth bottleneck. To make our comparisons fair with Clockwork, we omit transferring input data for Clockwork.

Shepherd is not open-sourced. We have communicated with Shepherd's authors to clarify the ambiguity in their pseudocode and implemented its Flex scheduling algorithm.

We used multiple workloads in our evaluations. Workloads differ in the following dimensions: (a) batching profiles of the models, (b) latency SLOs, (c) invocation popularity across models, (d) request arrival patterns, and (e) changes in the average request rate. The arrival of the request follows the Poisson distribution unless specified.

Our evaluation answers the following questions:
\squishlist
    \item How is the scheduling quality of Symphony under different kinds of workloads?
    \item How much GPUs cost can Symphony save?
    \item Where do performance benefits come from?
    \item Can Symphony work with cluster auto-scaling tools?
    \item Will the centralized scheduler become a bottleneck?
    \item What is the impact of RDMA?
    \item Does Symphony respond to fast-changing workloads?
\squishend

\subsection{End-to-end Goodput}

\begin{figure}[t]
\centering
\includegraphics[width=\linewidth]{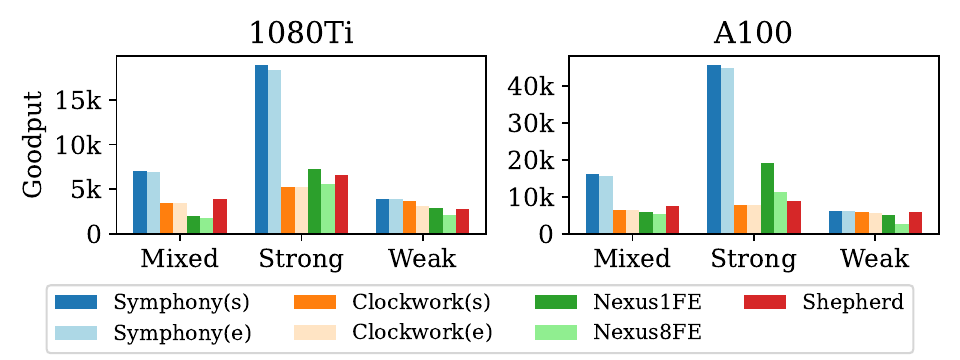}
\caption{Goodput comparison in mixed-model settings}
\label{fig:mixed-model}
\end{figure}

We collected a mixed model zoo consisting of 37 widely-used DNN models, including variants of DenseNet~\cite{densenet}, EfficientNet~\cite{efficientnet,efficientnetv2}, Inception~\cite{inceptionv3,inceptionresnet}, MobileNet~\cite{mobilenet,mobilenetv2,mobilenetv3}, NASNet~\cite{nasnet}, ResNet~\cite{resnet,resnetv2}, VGG~\cite{vgg}, Xception~\cite{xceptionnet}, SSDMobileNet~\cite{liu2016ssd}, and Bert~\cite{devlin2018bert}. These models span a wide range of parameter sizes and execution speeds. Latency SLOs of models vary from 20 to 400 milliseconds. Batching characteristics of each model also differ. The full list of models and latency SLO settings is available in Appendix~\ref{app:modelzoo}.

We evaluated Symphony, Clockwork, Nexus, and Shepherd on the mixed model zoo with 64 emulated GPUs in two separate clusters with NVIDIA 1080Ti and A100, respectively. Because both Symphony and Clockwork take the centralized approach, we ran both systems with both scheduler-only (s) and end-to-end (e) configurations. We also run Shepherd in the scheduler-only setting. Scheduler-only runs only the scheduler, and the load generator sends requests in the same process. The end-to-end configuration runs the scheduler, frontends, backends, and load generators on separate machines. Since Nexus frontends take part in scheduling, we run Nexus with a single frontend and eight frontends, respectively, to observe the overhead of distributed scheduling.

Based on batching characteristics, we run the experiment in three settings: \emph{Mixed} runs all models listed in the model zoo; \emph{Strong} runs models whose $\beta/\alpha > 2$, i.e., models that have strong batching effect; \emph{Weak} runs models whose $\beta/\alpha < 2$, i.e., models that do not benefit much from batching.

The goodput of each system is shown in Figure~\ref{fig:mixed-model}. Comparing the goodput between (s) and (e), we can see that both Symphony's and Clockwork's schedulers have good control over the cluster and are able to predict the end-to-end performance. Comparing Nexus1FE and Nexus8FE, we observe 11\% to 45\% goodput loss from distributed scheduling.

When running all models, Symphony shows 2.0-2.4x the goodput of the baselines. When running models with strong batch effects, Symphony shows 3.5x the goodput of the baseline systems for 1080Ti and 5.7x for A100, indicating that Symphony achieves better scheduling quality. It is worth pointing out that the advantage of Symphony is less prominent when running models with little batching effect. Symphony goodput is 23\% and 10\% higher in the 1080Ti cluster and the A100 cluster, respectively, in the \emph{Weak} setup.

\subsection{Savings in GPUs}

\begin{figure}[t]
\centering
\includegraphics[width=\linewidth]{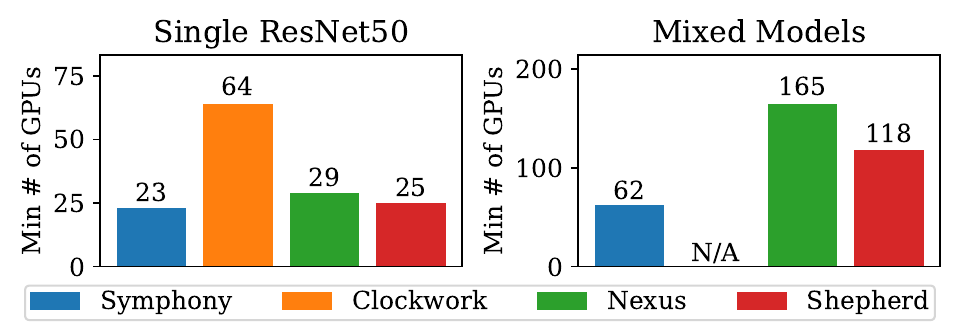}
\caption{Minimum number of GPUs required for 15k RPS}
\label{fig:gpu-saving}
\end{figure}

We compare the four systems on the minimum number of GPUs required for serving 15k RPS as depicted in~\Cref{fig:gpu-saving}.
We examine the following two workloads: a single ResNet50 model with 25 ms latency SLO, and all 37 mixed models with various latency SLO in~\Cref{tab:full-latency-profiles-a100}.
In this experiment, we use an emulated A100 cluster.
In the single-model case, Symphony saves 2 to 6 GPUs compared with Shepherd and Nexus. Clockwork requires more than double the GPUs of the other three systems since it does not explicitly optimize for batching efficiency.
Symphony's resource consolidation effect is larger for the mixed-model cases because it is more challenging for baseline systems to achieve large batch sizes when serving multiple models.
To deliver the same goodput, Nexus and Shepherd need 166\% and 90\% more GPUs.
Clockwork cannot deliver the desired goodput because its scheduling algorithm has a very high time complexity.

\subsection{Scheduling Quality}
\label{sec:eval-scheduling-quality}

We test Symphony's scheduling quality under various workload characteristics. Then, using two setups as examples, we take a closer examination to understand why Symphony achieves better scheduling quality.

\begin{figure}[t]
\centering
\includegraphics[width=\linewidth]{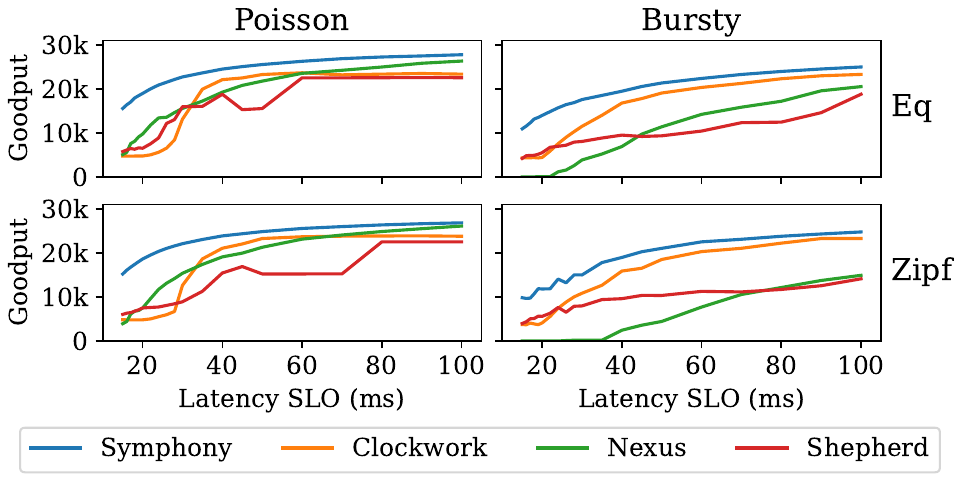}
\caption{Effect of workload characteristics on goodput}
\label{fig:slo}
\end{figure}

\paragraph{Varying the workload characteristics.}
We conducted experiments with changes in three dimensions of workloads: latency SLO, model popularity, and request arrival. The experiments use 20 models whose batching profile is similar to ResNet50; this would represent specialized variants of the model for different applications~\cite{Raffel2019ExploringTL,Howard2018UniversalLM,Pan2010ASO}. All models are set to the same latency SLO, varying from 15ms to 100ms across data points. Popularity among models has two options: equally popular or Zipfian distribution with a shape of 0.9. The arrival process is either a Poisson or Gamma distribution with shape 0.05 (i.e., bursty request rate). 32 emulated GPUs are used in the experiments.

Figure~\ref{fig:slo} shows the goodput of the different systems under the various combinations of the experimental setups. Our findings include: 1) Symphony provides significant benefits in the tight-SLO cases across all four setups, and this is because deferred batch scheduling optimizes batch efficiency. 2) Nexus does not work well with bursty workloads because of the static partitioning of models to backends; it does not enjoy the statistical multiplexing benefits of the other systems. 3) Loose-SLO cases are less challenging, with all systems producing relatively good schedules. This is because the batch sizes are large; therefore, the marginal goodput improvements from larger batch sizes are small.

\begin{table*}[tb]
\centering
{\small
\begin{tabular}{|c|c|c|c|c|cr|cr|crrr|}
\hline
\multirow{2}{*}{GPUs} & \multirow{2}{*}{Model} & \multirow{2}{*}{$\alpha$(ms)}  & \multirow{2}{*}{$\beta$(ms)}    & \multirow{2}{*}{SLO}    & \multicolumn{2}{c|}{No Coordination}                       & \multicolumn{2}{c|}{Staggered}                       & \multicolumn{4}{c|}{Measured Goodput (r/s)}                                                                \\ \cline{6-13} 
                       & &                            &                             &                         & \multicolumn{1}{c|}{BS} & \multicolumn{1}{r|}{Tpt } & \multicolumn{1}{c|}{BS} & \multicolumn{1}{r|}{Tpt} & \multicolumn{1}{c|}{Symphony} & \multicolumn{1}{c|}{Clockwork} & \multicolumn{1}{c|}{Nexus}  & \multicolumn{1}{c|}{Shepherd} \\ \hline
8 & \cite{resnet}          & \multicolumn{1}{r|}{1.053} & \multicolumn{1}{r|}{5.072}  & \multicolumn{1}{r|}{25ms} & \multicolumn{1}{r|}{7}  & 4501 r/s                    & \multicolumn{1}{r|}{16} & 5839 r/s                   & \multicolumn{1}{r|}{5264}     & \multicolumn{1}{r|}{1358}      & \multicolumn{1}{r|}{4027} & 4445                       \\ \hline
8 & \cite{inceptionresnet} & \multicolumn{1}{r|}{5.090} & \multicolumn{1}{r|}{18.368} & \multicolumn{1}{r|}{70ms} & \multicolumn{1}{r|}{3}  & 713 r/s                    & \multicolumn{1}{r|}{8}  & 1083 r/s                    & \multicolumn{1}{r|}{926}      & \multicolumn{1}{r|}{458}       & \multicolumn{1}{r|}{618} & 778                        \\ \hline
\end{tabular}
}
\caption{Theoretical analysis of batching and empirical measurement of goodput. (BS: Batch Size, Tpt: Throughput)}
\label{tab:stagger-exec}
\end{table*}

\paragraph{Batch Size.}
\Cref{sec:schedule-example-stagger} discussed the staggered execution pattern. The worst queuing delay in staggered execution is $\nicefrac{\ell(b)}{N}$. For Nexus, due to its distributed design, the worst queuing delay for a request is $\ell(b)$. Assuming requests arrive at a constant gap, the analytical solution of batch size can be calculated as
$\left\lfloor\left(\nicefrac{\mathrm{SLO}}{2}-\beta\right)/\alpha\right\rfloor$ and 
$\left\lfloor\left(\nicefrac{\mathrm{SLO}}{(1+\nicefrac{1}{N})}-\beta\right)/\alpha\right\rfloor$, 
respectively. Given batch size $b$, the throughput of $N$ GPUs can be calculated as $Nb / (\alpha b + \beta)$.

To understand how batch size affects goodput and to study how close is the batching behavior to the ideal staggered execution, we run a single copy of ResNet50~\cite{resnet} and InceptionResNetV2~\cite{inceptionresnet} separately with 8 GPUs on Symphony, Clockwork, and Nexus. Requests arrive in a Poisson distribution. The details of the models and the goodput on each system are listed in Table~\ref{tab:stagger-exec}, and the distribution of batch size is shown in Figure~\ref{fig:stagger}.

Table~\ref{tab:stagger-exec} shows the analytical batch size and the throughputs for the different scheduling approaches. The data shows that the staggered execution can run twice the batch size and achieve $30\%\sim50\%$ higher throughput compared to execution without coordination. The empirical measurement shows that the goodput of Symphony is close to the analytical throughput of staggered execution, and the goodput of Nexus is close to the analytical throughput of execution without coordination. The difference between the real system and analytical calculation comes from differences in request arrival patterns and delays in real systems.

Figure~\ref{fig:stagger} shows the batch size distribution. When running ResNet50, the majority of requests run with batch sizes greater than 14 and 6 for Symphony and Nexus, respectively. When running InceptionResNetV2, most batch sizes are greater than 8 and 3, respectively. The observed batch sizes is close to the analytical solutions. Thus, Symphony's scheduler generates batches whose sizes are close to that of the optimal staggered execution and thereby achieves high goodput.
Clockwork has an extremely low batch size since its scheduling algorithm does not consider batching efficiency.
Shepherd runs larger batches than Nexus and achieves higher throughput thanks to its centralized scheduler and preemption.
Compared to Symphony, Shepherd's batch size distribution is more spread out, indicating that Shepherd cannot form the ideal staggered execution due to its eager scheduling, hence we can see that Shepherd achieves a much lower goodput in \Cref{tab:stagger-exec}.

\begin{figure}[t]
\centering
\includegraphics[width=\linewidth]{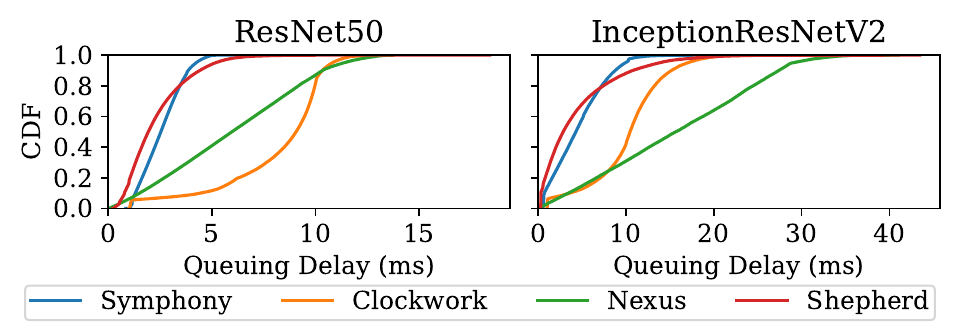}
\caption{Queuing Delay}
\label{fig:queue-delay}
\end{figure}

\paragraph{Queuing Delay.}
Figure~\ref{fig:queue-delay} shows the request queuing delay, which is defined as the duration starting from the system receiving the request to a GPU initiating a batch containing the given request. Symphony's queuing delay is 2x to 3x shorter than Nexus and Clockwork, allowing more of the SLO budget to be spent on execution. The longest queuing delay in Nexus is around half the latency SLO due to its lack of coordination. Although Clockwork uses a centralized scheduler, its longest queuing delay does not improve over Nexus in the ResNet50 case. Shepherd's queuing delay is comparable to Symphony, but it does not translate to larger batch sizes, as analyzed in the batch size evaluation.

\subsection{Auto Scaling}
\label{sec:eval-auto-scaling}
We study whether our system and baseline systems can provide robust signals to cluster auto-scaling tools. We use 10 ResNet models with 100ms latency SLO with 24 emulated GPUs in the experiment.

\paragraph{Stability.}
Ideally, a model serving system should provide a stable goodput even when the offered request rate exceeds the cluster's maximum capacity. 
Figure~\ref{fig:stability-gpu-cycle} (Left) shows that both Symphony and Nexus have flat-top behavior because both systems try to maintain a desirable batch size.
Shepherd's goodput shows some fluctuation when overloaded.
As shown in the graph, Clockwork's goodput starts to degrade as soon as the system is overloaded. This unstable goodput makes it hard to perform autoscaling.

\paragraph{GPU Cluster Utilization.}
Figure~\ref{fig:stability-gpu-cycle} (Right) measures cluster utilization as the average fraction of time that GPUs are busy. Clockwork, Nexus, and Shepherd reach full GPU busy levels before reaching their peak goodputs, indicating that both systems run with suboptimal batches. Symphony's GPU utilization gradually increases and reaches full GPU busy level at near-peak goodput, which means that Symphony can efficiently consolidate GPU usage. Hence, GPU utilization of Symphony is a good signal to cluster autoscaling tools.

\subsection{Scalability}\label{sec:eval-scalability}

\begin{figure}[tb]
\centering
\includegraphics[width=\linewidth]{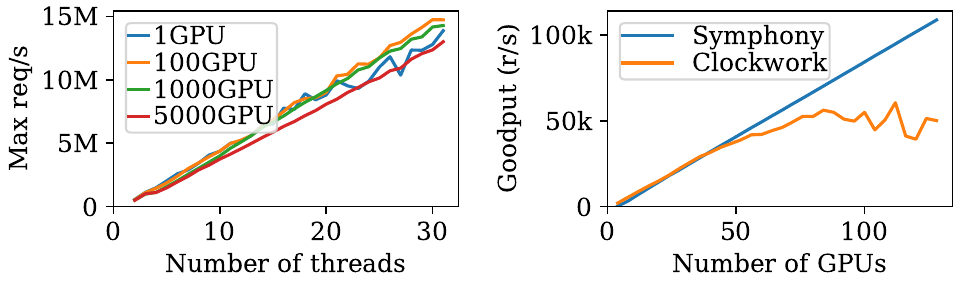}
\caption{(Left) Symphony scheduler multicore scalabilty. (Right) Goodput varying the number of GPUs.}
\label{fig:scalability}
\end{figure}

Since the centralized scheduler is involved in all requests, we want to ensure it can keep up with the request rates.
Figure~\ref{fig:scalability} (Left) measures the maximum request rate the Symphony scheduler can handle on a 32-core AMD EPYC 7502 CPU. We run this benchmark with the scheduler alone, without sending network messages or running GPUs. Requests and GPUs in these benchmarks are in-process objects used by the scheduler.
Since there is no information exchange between \textsc{ModelThread}s, we observe that the throughput increases linearly with the number of threads. The plot also proves that the single-threaded \textsc{RankThread} is not a bottleneck. Although \textsc{ModelThread}s produces Candidates at line rate, \textsc{RankThread} only needs to pick up the latest candidate from \textsc{ModelThread}s. Therefore, a single \textsc{RankThread} can support dozens of \textsc{ModelThread}s.
The plot also shows that adding more GPUs does not affect the scalability, as we use advanced data structures to manage states.

Figure~\ref{fig:scalability} (Right) serves 20 equally popular ResNet-like models with 100ms SLO. Symphony's goodput increases linearly with the number of emulated GPUs. 
On the other hand, Clockwork was not designed for multicore scalability, and its use of locks on critical paths limits the overall throughput.
We omit the comparison with Shepherd due to the lack of a canonically optimized implementation.

\subsection{Impact of network}
\label{sec:eval-tail-latency}

\begin{figure}[t]
\centering
\includegraphics[width=\linewidth]{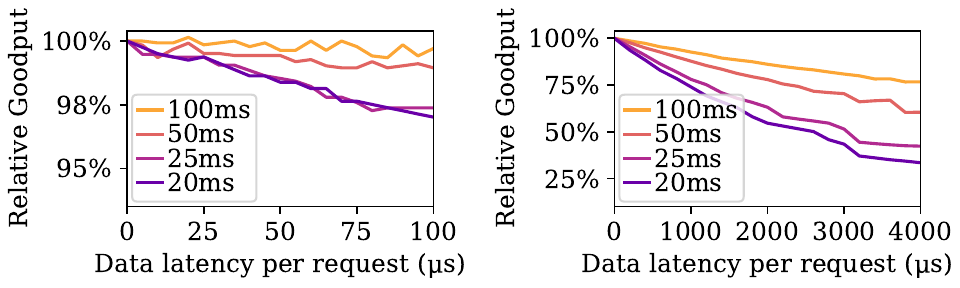}
\caption{Effect of network latency on model serving goodput. (Left) within RDMA range. (Right) within TCP range.}
\label{fig:ddata}
\end{figure}

This subsection quantifies the benefits of using RDMA-based communications in Symphony. The network-level benchmarks on TCP and RDMA are in Appendix~\ref{app:testbed}. The lowest latency and 99.99-th tail in the RDMA network are close, at 24\textmu{}s and 33\textmu{}s respectively. The median latency of TCP is 3034\textmu{}s, and its 99.99-th tail is 12x the median.
To study how network latency affects the model serving goodput, we run a workload consisting of 20 evenly popular models of similar batching profiles on an emulated 32-GPU cluster with four settings of latency SLO: 20ms, 25ms, 50ms, and 100ms. Figure~\ref{fig:ddata} (Left) shows that the network latency within the RDMA range only marginally reduces goodput. Figure~\ref{fig:ddata} (Right) shows that network latency within the TCP tail latency range significantly hurts goodput by up to 70\%. We observe that the tighter the latency SLO is, the more severely the goodput is impacted by network latency.
This is because a backend cannot fetch input data from frontends until the batch metadata is sent by the scheduler.
The scheduler always uses the high percentile bound of network latency as the network delay estimation and would have to make earlier dispatch decisions. Therefore, unpredictable and high latencies lead to a significant decrease in the goodput.

\subsection{Large cluster and changing workload}

Lastly, we evaluate our system in a large cluster setup with a changing workload. The emulated cluster consists of 512 GPUs. The workload consists of 24 models with different batching characteristics and different SLOs. The request rate for each model is synthesized from 150 hours of videos. 

\begin{figure}[ht]
\centering
\includegraphics[width=\linewidth]{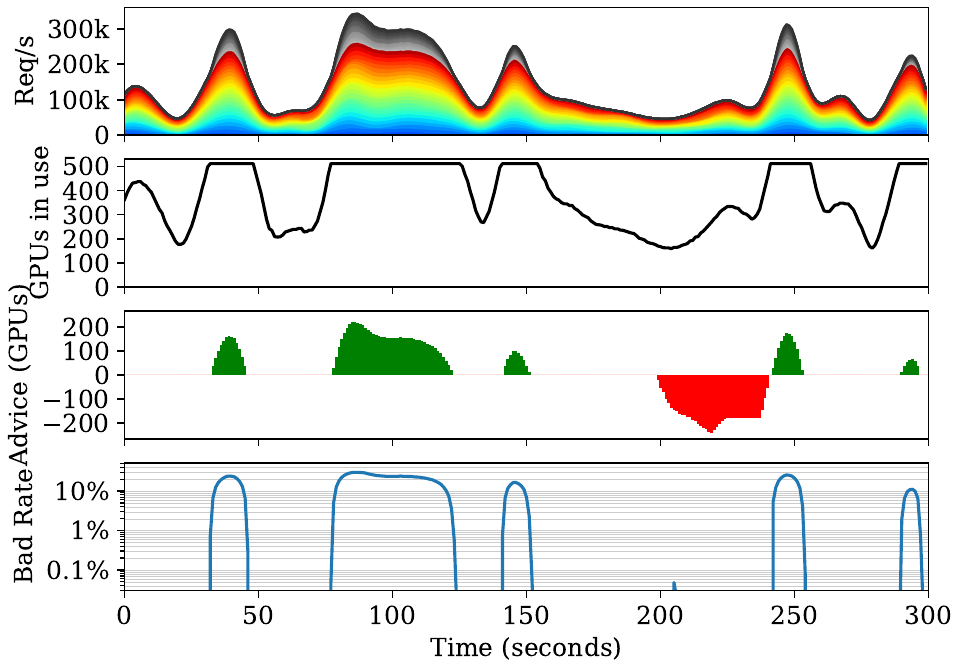}
\caption{A changing workload on a 512-GPU cluster}
\label{fig:replay}
\end{figure}

Figure~\ref{fig:replay} shows the result. Color in the first subplot represents the goodput for each model. Grayscale in the first subplot shows the unsatisfied requests for each model. The stacked graph shows the overall request rate per second in total. The second subplot shows the number of GPUs used in the system. Symphony can consolidate GPU usage when the cluster is underloaded. The third subplot shows Symphony's advice to the cluster's autoscaling system regarding adding or removing GPUs. 
The fourth subplot shows the bad rate. Symphony can maintain a low bad rate when the cluster is not overloaded.

%% file: tex/90-conclusion.tex
\section{Discussion}

\paragraph{Failure Recovery}
The centralized scheduler is a single point of failure.
We provide a millisecond-scale fast recovery design.
We set aside a few hot-spare schedulers in the cluster whose only task is to have frequent heartbeat exchanges with the primary.
With the help of RDMA, a failure can be detected in a few hundred microseconds.
One of the hot spares can quickly switch to the primary.
The new scheduler can query backends to populate GPU work states and start to accept new requests immediately.
Unfinished requests during failure are inevitably lost and can be retried by clients.

\paragraph{Generalization to Large Language Models} Our system is optimized for DNN inference with stringent latency SLOs. For instance, the machine learning models we focus on, such as ResNet, are widely used in computer vision and video analytics. In these domains, ML inference usually requires tight SLOs for timely object detection and real-time vehicle tracking purposes. Large language models (LLMs)~\cite{attention} are an emerging type of DNN with a very different performance requirement. For improving LLM inference, batching is also important~\cite{orca,PagedAttention} but must be considered with other factors, including how to improve kernel operator efficiency~\cite{FlashAttention,awq,gptq,smoothquant,llm-int8}, how to partition model weights to accelerators through model parallelism~\cite{alpa,DeepSpeed-Inference,megatron-lm}, and how to maintain the KV cache across multiple model invocations. The performance requirements of LLMs are also quite different. An LLM's per-token latency is usually around dozens of ms, and the token decoding latency has other factors including the window length. How to extend our deferred batch scheduling idea to LLMs is an interesting future work.

\section{Related work}

\paragraph{Model Serving Systems:} 
Earlier, we described Symphony's relationship to Nexus~\cite{nexus}, Clockwork~\cite{clockwork}, Shepherd~\cite{shepherd}, Clipper~\cite{Clipper}, and Tensorflow Serving~\cite{TFServing}. In addition, Salus~\cite{salus} is a system that emphasizes loading multiple models into the GPU memory, but it mainly focuses on training and only discusses inference on a single-machine setup. INFaaS~\cite{INFaaS} focuses on model selection, trading off between the inference latency and the DNN model accuracy, which is in a different scope from ours.
Perseus~\cite{Perseus-arxiv} is a good motivational paper for multi-tenant model serving, but they didn't dive deep into scheduling algorithms. Our paper proposes a solution in this direction.
Pretzel~\cite{lee2018pretzel}'s most important technique is sharing layers across models. We believe that Symphony's techniques can be extended to batch invocations of shared layers as well.
MArk~\cite{mark-atc19} and Morphling~\cite{morphling-socc21} focus more on autoscaling and machine provisioning. Our scheduler's statistics and autoscaling advice can be good signals to these cluster management tools.
Llama~\cite{Llama-socc21} and Scrooge~\cite{Scrooge-socc21} focus more on complex query pipelines whereas we focus on the batching efficiency of individual models. These systems can adopt our techniques to further improve the efficiency of individual models in a pipeline.
MArk, Morphling, and Llama try to cut down costs on cloud infrastructures by choosing the right VMs/accelerators; our work is orthogonal and aims at improving the efficiency of the chosen target.

\paragraph{Load Balancing:} Load balancing has been studied extensively, and many algorithms have been proposed to maximize throughput and minimize response time. There are two categories of algorithms: static load balancing algorithms (e.g., consistent hashing~\cite{karger1997consistent}, and CFS~\cite{dabek2001wide}) distribute requests based on the prior knowledge of the capability of a node, while dynamic load balancing algorithms (e.g., the power of two choices~\cite{mitzenmacher2001power}, Distcache~\cite{liu2019distcache}, and Sparrow~\cite{Sparrow}) take account into the run-time properties collected of nodes. Slicer~\cite{adya2016slicer} is a production-grade auto-sharder. However, Slicer does not apply to our use case because we have a tighter bound of latency SLO, and by Slicer's distributed nature, it will result in a less balanced load.

\section{Conclusion}

We present the design of Symphony, an optimized DNN model serving system for GPU clusters. Symphony uses a novel \emph{deferred batch scheduling} algorithm that improves batch sizes while meeting latency SLO. The number of GPUs that Symphony uses is \emph{proportional to the load}, enabling easy autoscaling. To support the algorithm, we craft a low-latency, multicore-scalable centralized scheduler capable of handling millions of requests per second and managing thousands of GPUs. Compared to prior systems, Symphony achieves 5x higher goodput when given the same number of GPUs and saves 60\% GPUs when serving the same workload.
Symphony's source code is available at \href{https://github.com/abcdabcd987/nexuslb}{\texttt{\nolinkurl{https://github.com/abcdabcd987/nexuslb}}}.

%% file: tex/99-appendix.tex
\newpage
\appendix

\section{Global Partitioning Algorithm}
\label{app:partition}

We first present details on the MILP optimization formulation that we use to partition the models across sub-clusters and clusters. We then provide an evaluation of the quality of approximated solutions to the MILP in the context of our system.

\subsection{Model formulation}

We want to partition $m$ models into $l$ sub-clusters. For each sub-cluster, the maximum dispatcher capability of request rate is $R_{\text{max}}$, and the maximum memory size of backends is $S_{\text{max}}$. Let $x_{ij}$ represents whether model $i$ is assigned to sub-cluster $j$. The request rate, static memory size, dynamic (runtime) memory size of model $i$ are $r_i$, $s_i$, $d_i$ respectively. The average request rate per sub-cluster is $\bar{R} = \sum_{i} r_i/l$. Similarly, the average static memory size per sub-cluster is $\bar{S}=\sum_{i} s_i/l$. The partitioning problem can be formulated as follows.
\begin{alignat}{3}
	 & \text{minimize} \quad   &  & \Delta{R}                                              + w\Delta{S}  \label{lp1}                                                                           \\
	 & \text{subject to} \quad &  & \sum\nolimits_{i} r_i x_{ij}                                                               \le R_{\text{max}}          &  & \quad \forall{j} \label{lp4}   \\
	 &                         &  & \sum\nolimits_{i} s_i x_{ij} + \max\nolimits_{i} d_ix_{ij}                                          \le S_{\text{max}} &  & \quad \forall{j} \label{lp5}   \\
	 &                         &  & \Big \lvert \sum\nolimits_{i} r_i x_{ij} - \bar{R} \Big \rvert                             \le \Delta{R}               &  & \quad \forall{j} \label{lp6}   \\
	 &                         &  & \Big \lvert \sum\nolimits_{i} s_i x_{ij} - \bar{S} \Big \rvert                             \le \Delta{S}               &  & \quad \forall{j} \label{lp7}   \\
	 &                         &  & \sum\nolimits_{j} x_{ij}                                                                   = 1                         &  & \quad \forall{i} \label{lp2}   \\
	 &                         &  & x_{ij}                                                                            \in \{0, 1\}                         &  & \quad \forall{i,j} \label{lp3}
\end{alignat}
(\ref{lp2}) and (\ref{lp3}) represent that each model is assigned to exactly one sub-cluster. (\ref{lp4}) requires that the request rate of each sub-cluster is less than the maximum dispatcher capability. (\ref{lp5}) requires that the static memory size plus the maximum dynamic (runtime) memory size is less than the maximum memory size of backends. (\ref{lp6}) and (\ref{lp7}) denote that each sub-cluster's request rate and static memory size should be close to the averages. Therefore, our objective (\ref{lp1}) is to minimize the differences to the averages as much as possible. The parameter $w$ is a coefficient that balances the request rate and model memory size.

\paragraph{Disruption Minimization}
The above model can generate an initial assignment when the system starts. When the system is running, we would like to generate the next assignment considering the current assignment, so the disruption is minimized.

Given the current assignment matrix $X'$ where $x'_{ij}$ denotes whether model $i$ is currently put into sub-cluster $j$. Let $c_{ij}$ denote the cost for loading/unloading model $i$ to/from sub-cluster $j$ (assuming the cost is symmetric between loading and unloading). The maximum allowed cost is $C_\text{max}$. Let $y_{ij}$ represent whether there is a change for model $i$ and sub-cluster $j$ as $y_{ij} = |x_{ij} - x'_{ij}|$. To bound the total cost of change within $C_\text{max}$, we have
\begin{equation}
    \sum\nolimits_{i} \sum\nolimits_{j} c_{ij}y_{ij} \leq C_\text{max} \, . \label{eq:c_change}
\end{equation}

\paragraph{Practical Implementation}
As solving general MILP problems is NP-hard, an optimal solution of the above model can not be obtained within a reasonable amount of time. Therefore, we need to use an approximated solution obtained within a time bound. 
Based on our testing with CPLEX solver and a running time of 10 seconds for $m = 400$ and $l = 4$, an approximate solution is significantly better than any randomly generated assignments.

\subsection{Effectiveness of Cluster Partitioning}

We next evaluate the effectiveness of the MILP partitioning model for partitioning the models and the workload across sub-clusters or clusters. We created a random solver as a baseline. The random solver will simply generate a random partition, check it against all the constraints, evaluate using the MILP objective, and repeat. It will output the partition with the least objective that satisfies all the constraints within a time constraint. Similarly, we use a solver to tackle the MILP problem and impose the same time limit for the solver (which is 10 seconds in our current system) to get the best solution so far. 

\begin{figure}[htb]
\centering
\includegraphics[width=\linewidth]{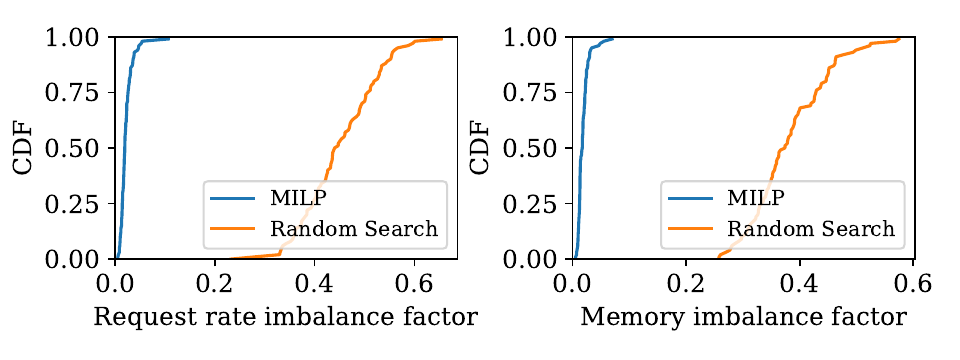}
\caption{Evaluating the effectiveness of MILP search for the partitioning problem}
\label{fig:partition}
\end{figure}

We generate multiple configurations to represent different scenarios. Each configuration contains a random selection of models from our model zoo, where we generate multiple variants as specialized instantiations of a given model. We consider large-scale partitioning problems that partitions 800 models across 20 partitions.
The request rate for each model is assumed to be independent and draw from an exponential distribution. We measure the goodness of a partition based on the imbalance factors for memory and load. The imbalance factor is defined as $\frac{\max - \min}{\text{avg}}$ across the different partitions generated by the algorithm. We compute the imbalance factor for both static memory and request rate across all dispatchers / subclusters. The smaller the imbalance metric, the better is the partition. Figure~\ref{fig:partition} depicts the CDF of the imbalance factors for memory and load across the different experiments. We observe that our MILP-based algorithm provides load-balanced partitions within our time window of 10 seconds, which makes it appropriate for epoch-level reconfigurations that happen every few minutes.

\section{Networking Performance of Our Testbed}
\label{app:testbed}

\begin{figure}[t]
\centering
\includegraphics[width=\linewidth]{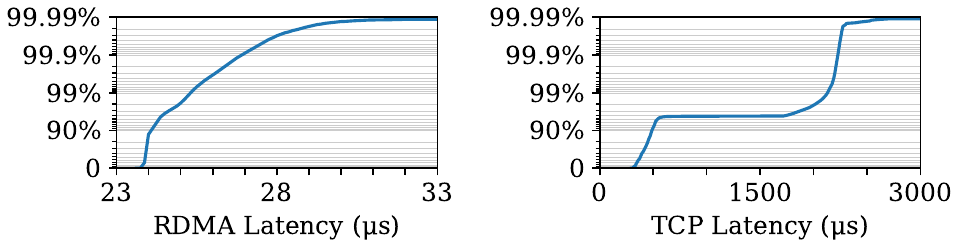}
\caption{RDMA and TCP tail latency}
\label{fig:tail-latency-rdma-vs-tcp}
\end{figure}

We study the performance difference between RDMA and TCP. We ran a network incast benchmark among eight servers. The incast behavior is a good representation of sending inputs of a batch from frontends to the executing backend. The benchmark concurrently reads eight objects of 150KB from each server. RDMA tail latency was measured on 56Gbps Infiniband. TCP tail latency was measured on 40Gbps Ethernet.

Figure~\ref{fig:tail-latency-rdma-vs-tcp} shows the incast benchmark result. The lowest RDMA latency is within 24 \textmu{}s, which is very close to the theoretical lower bound of 21.5 \textmu{}s. The 99.99-th percentile latency in the RDMA network is within 33 \textmu{}s. Thus RDMA network is both low-latency and highly predictable. TCP, on the other hand, is not only slower but also has a very long tail. The 99.99-th percentile latency is 12x the median.

\section{Model Zoo Details}
\label{app:modelzoo}
Table~\ref{tab:full-latency-profiles-1080Ti} and Table~\ref{tab:full-latency-profiles-a100} show the profiles of models used in the evaluation section for 1080Ti and A100 respectively. Latency SLO associated with each model ensures that each model can run with batch size greater than or equal to 4.

\begin{table*}[t]
\parbox{0.48\linewidth}{
\centering
\begin{tabular}{|l|r|r|r|r|}
\hline
Name              & $\alpha$ (ms)  & $\beta$ (ms) & $\beta/\alpha$    & SLO\\
\hline
NASNetMobile      & 0.570  & 14.348 & 25.172 & 33ms  \\
MobileNetV3Small  & 0.335  & 5.350  & 15.970 & 20ms  \\
DenseNet169       & 1.271  & 13.618 & 10.714 & 37ms  \\
DenseNet121       & 1.061  & 10.312 & 9.719  & 29ms  \\
DenseNet201       & 1.733  & 15.687 & 9.052  & 45ms  \\
EfficientNetV2B0  & 1.006  & 7.493  & 7.448  & 23ms  \\
MobileNetV3Large  & 0.820  & 5.256  & 6.410  & 20ms  \\
InceptionV3       & 1.964  & 8.771  & 4.466  & 33ms  \\
EfficientNetV2B1  & 1.661  & 7.247  & 4.363  & 27ms  \\
ResNet50V2        & 1.409  & 5.947  & 4.221  & 23ms  \\
ResNet152V2       & 3.471  & 13.049 & 3.759  & 53ms  \\
ResNet101V2       & 2.438  & 9.095  & 3.731  & 37ms  \\
InceptionResNetV2 & 5.090  & 18.368 & 3.609  & 77ms  \\
EfficientNetB0    & 1.569  & 5.586  & 3.560  & 23ms  \\
MobileNetV2       & 1.180  & 3.483  & 2.952  & 20ms  \\
ResNet101         & 3.164  & 9.065  & 2.865  & 43ms  \\
EfficientNetB1    & 2.489  & 6.674  & 2.681  & 33ms  \\
ResNet50          & 2.050  & 5.378  & 2.623  & 27ms  \\
EfficientNetV2B2  & 2.254  & 5.896  & 2.616  & 29ms  \\
VGG19             & 3.059  & 7.857  & 2.568  & 40ms  \\
ResNet152         & 4.599  & 11.212 & 2.438  & 59ms  \\
MobileNet         & 1.009  & 2.390  & 2.369  & 20ms  \\
VGG16             & 2.734  & 5.786  & 2.116  & 33ms  \\
EfficientNetB2    & 3.446  & 5.333  & 1.548  & 38ms  \\
EfficientNetV2B3  & 4.072  & 5.981  & 1.469  & 44ms  \\
NASNetLarge       & 17.656 & 18.952 & 1.073  & 179ms \\
EfficientNetV2S   & 8.463  & 8.862  & 1.047  & 85ms  \\
EfficientNetB3    & 5.924  & 4.849  & 0.819  & 57ms  \\
EfficientNetV2L   & 40.313 & 28.208 & 0.700  & 378ms \\
EfficientNetV2M   & 22.619 & 14.786 & 0.654  & 210ms \\
EfficientNetB5    & 23.435 & 10.301 & 0.440  & 208ms \\
Xception          & 4.751  & 2.046  & 0.431  & 42ms  \\
SSDMobilenet      & 23.778 & 9.729  & 0.409  & 209ms \\
EfficientNetB4    & 12.088 & 4.412  & 0.365  & 105ms \\
BERT              & 7.008  & 0.159  & 0.023  & 56ms \\
\hline
\end{tabular}
\caption{Model profiles on an NVIDIA 1080Ti.}
\label{tab:full-latency-profiles-1080Ti}
}
\hfill
\parbox{0.48\linewidth}{
\centering
\begin{tabular}{|l|r|r|r|r|}
\hline
Name              & $\alpha$ (ms)  & $\beta$ (ms) & $\beta/\alpha$    & SLO\\
\hline
DenseNet121       & 0.054  & 10.546    & 195.296 & 21ms  \\
DenseNet201       & 0.304  & 14.345    & 47.188  & 31ms  \\
DenseNet169       & 0.289  & 13.365    & 46.246  & 29ms  \\
ResNet50V2        & 0.135  & 5.560     & 41.185  & 20ms  \\
EfficientNetB0    & 0.115  & 4.326     & 37.617  & 20ms  \\
ResNet101         & 0.284  & 8.266     & 29.106  & 20ms  \\
ResNet152         & 0.390  & 10.449    & 26.792  & 24ms  \\
ResNet101V2       & 0.391  & 8.219     & 21.020  & 20ms  \\
MobileNetV3Large  & 0.196  & 4.072     & 20.776  & 20ms  \\
EfficientNetB1    & 0.291  & 5.797     & 19.921  & 20ms  \\
ResNet50          & 0.268  & 5.172     & 19.299  & 20ms  \\
ResNet152V2       & 0.589  & 10.054    & 17.070  & 24ms  \\
MobileNetV2       & 0.190  & 2.892     & 15.221  & 20ms  \\
EfficientNetV2B3  & 0.543  & 7.596     & 13.989  & 20ms  \\
InceptionResNetV2 & 1.112  & 15.27     & 13.732  & 39ms  \\
EfficientNetV2B1  & 0.443  & 5.929     & 13.384  & 20ms  \\
NASNetMobile      & 0.536  & 6.860     & 12.799  & 20ms  \\
EfficientNetV2B0  & 0.377  & 4.272     & 11.332  & 20ms  \\
EfficientNetB2    & 0.520  & 5.333     & 10.256  & 20ms  \\
MobileNetV3Small  & 0.315  & 3.211     & 10.194  & 20ms  \\
InceptionV3       & 0.913  & 6.732     & 7.373   & 20ms  \\
MobileNet         & 0.285  & 1.901     & 6.670   & 20ms  \\
EfficientNetV2S   & 1.454  & 7.378     & 5.074   & 26ms  \\
EfficientNetV2B2  & 0.901  & 4.532     & 5.030   & 20ms  \\
VGG16             & 0.660  & 2.252     & 3.412   & 20ms  \\
EfficientNetB3    & 1.239  & 4.205     & 3.394   & 20ms  \\
Xception          & 0.801  & 2.638     & 3.293   & 20ms  \\
VGG19             & 0.893  & 2.181     & 2.442   & 20ms  \\
NASNetLarge       & 3.464  & 7.154     & 2.065   & 42ms  \\
EfficientNetV2M   & 4.479  & 6.861     & 1.532   & 49ms  \\
EfficientNetB4    & 2.881  & 4.103     & 1.424   & 31ms  \\
EfficientNetV2L   & 7.520  & 6.675     & 0.888   & 73ms  \\
EfficientNetB5    & 6.121  & 2.283     & 0.373   & 53ms  \\
SSDMobilenet      & 19.448 & 4.442     & 0.228   & 164ms \\
EfficientNetB6    & 9.754  & 1.984     & 0.203   & 82ms  \\
EfficientNetB7    & 16.339 & 2.751     & 0.168   & 136ms \\
BERT              & 7.353  & 0.222     & 0.030   & 59ms \\
\hline
\end{tabular}
\caption{Model profiles on an NVIDIA A100.}
\label{tab:full-latency-profiles-a100}
}
\end{table*}

\section{Scheduling Algorithm Pseudocode}
\label{appendix:pseudocode}

\Cref{code:rankmt} shows an extended version of \Cref{alg:frontrun-pseudocode}. The main differences are:
(1) This version accounts for network latency.
(2) This version includes the multicore scalability design.

\begin{figure*}[t]
\centering
\begin{minipage}{0.5\textwidth}
\begin{minted}{python}
struct Candidate: bs, exec_at, latest
def delay(bs): return d_ctrl + d_data * bs
class ModelThread:
  model: ModelID
  batch: BatchPolicy
  drop_timer: Timer
  c: Optional[Candidate]

  # Frontend -> ModelThread
  def schedule(req):
    batch.enqueue(req)
    update_candidate(-inf)
    rank_thread.inform_candidate(model, c)

  # RankThread -> ModelThread
  def granted_gpu(gpu, gpu_free_at):
    update_candidate(gpu_free_at)
    if c is None:
      free_at = max(now(), gpu_free_at)
    else:
      inputs = batch.pop_inputs()
      send(gpu, ExecutionMsg(inputs, c.exec_at))
      free_at = c.exec_at + exec_elapse(len(inputs))
      update_candidate(-inf)
    rank_thread.inform_gpu(gpu, free_at)
    rank_thread.inform_candidate(model, c)

  def update_candidate(gpu_free_at):
    batch.update_batch(gpu_free_at)
    bs = len(batch.inputs)
    if bs > 0:
      dl = batch.inputs[0].deadline
      drop_timer.cancel()
      drop_timer.set(dl, lambda: update_candidate(-inf))
      earliest = max(now()+delay(bs), gpu_free_at)
      frontrun_at = dl - exec_elapse(model, bs+1)
      exec_at = max(earliest, frontrun_at)
      latest = dl - exec_elapse(model, bs)
      c = Candidate(bs, exec_at, latest)
    else:
      c = None
    for req in batch.pop_dropped():
      send(req.frontend, QueryDroppedMsg(req))
\end{minted}
\end{minipage}%
\begin{minipage}{0.5\textwidth}
\begin{minted}{python}
class RankThread:
  gpu_free_at: Map[GpuID, TimePoint]
  gpu_timer: Map[GpuID, Timer]
  model_timer: Map[ModelID, Timer]
  mc: Map[ModelID, Candidate]

  # ModelThread -> RankThread
  def inform_candidate(m: ModelID, c: Optional[Candidate]):
    del mc[m]
    model_timer[m].cancel()
    if c is not None:
      model_timer[m].set(c.exec_at-delay(c.bs),
          lambda: on_model_timer(m, c))

  # ModelThread -> RankThread
  def inform_gpu(gpu: GpuID, free_at: TimePoint):
    gpu_free_at[gpu] = free_at
    set_gpu_timer()

  def on_model_timer(m: ModelID, c: Candidate):
    gpu, free_at = gpu_free_at.get_earliest()
    if free_at <= c.exec_at:
      gpu_free_at[gpu] = +inf
      model_thread[m].granted_gpu(gpu, free_at)
    else:
      mc[m] = c
      set_gpu_timer()
    
  def set_gpu_timer():
    gpu, free_at = gpu_free_at.get_earliest()
    gpu_timer[gpu].cancel()
    if len(mc) > 0:
      m, c = mc.get_by_max_delay()
      gpu_timer[gpu].set(free_at-delay(c.bs),
          lambda: on_gpu_timer(gpu))
  
  def on_gpu_timer(gpu: GpuID):
    free_at = gpu_free_at[gpu]
    Remove (m,c) from mc where free_at > c.latest
    if len(mc) > 0:
      m, c = mc.get_by_min_latest()
      model_thread[m].granted_gpu(gpu, free_at)
    set_gpu_timer()
\end{minted}
\end{minipage}
\caption{Pseudo-code of the scheduling algorithm.}
\label{code:rankmt}
\end{figure*}